\documentclass[letterpaper, 10 pt, conference]{ieeeconf}  

\IEEEoverridecommandlockouts                              

\usepackage{amsmath,amsfonts}
\usepackage[algo2e,ruled,vlined,linesnumbered,resetcount,norelsize]{algorithm2e}

\usepackage{array}
\usepackage[caption=false,font=normalsize,labelfont=sf,textfont=sf]{subfig}
\usepackage{textcomp}
\usepackage{stfloats}
\usepackage{float}
\usepackage{url}
\usepackage{bm}
\usepackage{verbatim}
\usepackage{graphicx}
\usepackage{cite}
\usepackage{mathtools} 
\usepackage{amssymb}  

\usepackage{amsthm}
\usepackage{algpseudocode}
\usepackage{xcolor}
\usepackage{accents}
\usepackage{mathrsfs}
\usepackage[normalem]{ulem}

\usepackage{soul}
\usepackage{booktabs}
\usepackage[table]{xcolor}
\usepackage{makecell}
\usepackage{datetime} 
\usepackage[useregional]{datetime2}

\usepackage{amsmath}        
\usepackage[ruled,vlined]{algorithm2e}

\usepackage{algorithm}
\usepackage{algpseudocode}
\usepackage[dvipsnames]{xcolor}
\usepackage{subcaption}
\usepackage{graphicx}
\usepackage{movement-arrows}
\usepackage{placeins}

\usepackage{multirow}

\usepackage[T1]{fontenc}
\usepackage[utf8]{inputenc}
\usepackage[skip=0.333\baselineskip]{caption}
\usepackage{siunitx} 
\newcolumntype{T}{S[table-format=3.3, input-symbols={()},
                    table-space-text-post={$^{***}$},
                    table-align-text-post=false]}
\usepackage[colorlinks=true, allcolors=black]{hyperref}
\usepackage{tabularx,booktabs}
\newcolumntype{C}{>{\centering\arraybackslash}X} 
\renewcommand{\baselinestretch}{.98}

\SetCommentSty{mycommfont}

\SetKwRepeat{Do}{do}{while}

\title{\LARGE \bf
Multi-robot Graph Traversal with Support Coordination under Stochastically Moving Adversaries
}

\author{Manshi Limbu, Xuan Wang, Gregory J. Stein, Daigo Shishika, and Xuesu Xiao 
\thanks{All authors are with George Mason University. {\tt\scriptsize \{klimbu2, xwang64, gjstein,  dshishik, xiao\}@gmu.edu}. 
\thanks{George Mason University.
{\tt\scriptsize \{klimbu2, xwang64, gjstein, dshishik, xiao\}@gmu.edu}.
}
}
}

\begin{document}
\maketitle
\thispagestyle{empty}
\pagestyle{empty}

\begin{abstract}
Cooperative multi-robot missions require team of robots to traverse environments where adversaries or hazards with stochastic dynamics induce time-varying traversal risk.
While support coordination---where robots assist teammates in traversing risky regions---can significantly reduce mission costs, its effectiveness depends on the team's ability to anticipate future risk.
We formulate support-based multi-robot graph traversal problem with stochastically moving adversaries, where future risky regions become uncertain as adversaries move through the environment.
When adversaries remain stationary, our formulation reduces to the static risky-edge setting. 
To address the stochastic case, we model individual adversaries as first-order Markov stay–move processes over graph edges and propagate their occupancy distributions over a finite planning horizon to obtain time-indexed edge-risk forecasts.
These forecasts inform the support candidate selection and joint robot path planning.
Experimental results show that forecast-informed support decisions consistently lower expected team cost relative to evaluated baselines in stochastic motion settings.

\end{abstract}

\section{Introduction}
\label{sec:introduction} 

Many multi-robot missions (e.g., search-and-rescue, convoy protection, reconnaissance, and patrolling) require team of robots to cooperatively navigate environments where traversal conditions evolve due to mobile adversaries or hazards whose motion follows stochastic dynamics. 
As these adversaries or hazards move, the set of risky regions changes, altering the cost of traversal across the environment. 
In these settings, \emph{support coordination}---where robots temporarily delay their own progress to assist or cover teammates traversing risky regions---can be a critical mechanism for reducing total team cost. 
By coordinating support between teammates, robots can safely traverse high-risk regions that would otherwise be expensive without teammate assistance~\cite{10341820}.

However, the effectiveness of such support coordination depends on the team’s ability to anticipate when and where risk will manifest. 
When risk is induced by \emph{adversaries with stochastic motion}, a planner that reasons only about the current risk configuration cannot exploit predictable temporal trends of future risk.
As a result, support decisions may be deployed too early, too late, or at locations that cease to be risky, leading to redundant waiting, detours, and increased total mission cost.
Effective support coordination in these settings, therefore, requires reasoning about how traversal risk evolves over time and where support is likely to be beneficial when that risk is encountered.

\begin{figure}[htbp]
    \centering
    \includegraphics[width=1.0\linewidth]{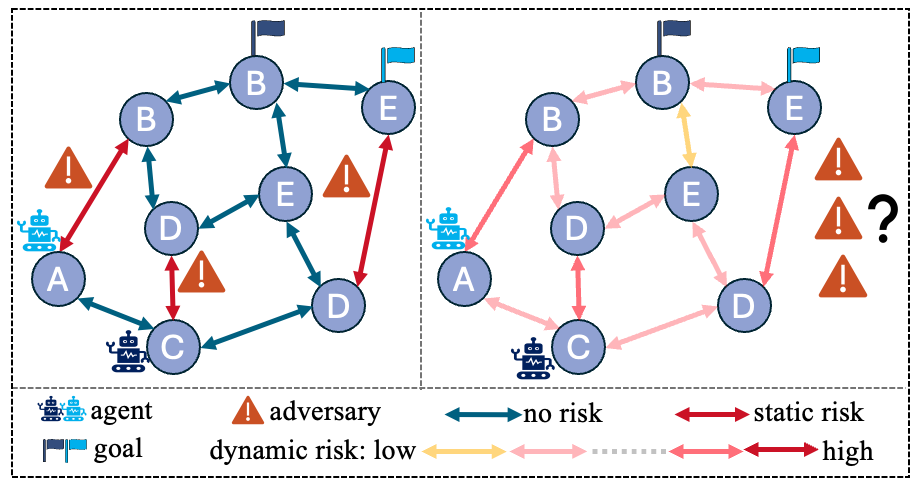} 
        \caption{\small From stationary to stochastic adversaries in multi-robot graph traversal. Stationary adversaries result in static edge risks (left), whereas adversaries with stochastic stay–move dynamics induce time-varying, dynamic edge risks (right), motivating anticipatory support coordination.}
    \label{fig:problem_figure}
\end{figure} 

Existing support-based frameworks~\cite{10341820,10341497,duggan2025uncertaintyawareplanningheterogeneousrobot} consider state-dependent traversal costs under environment or team conditions. TCGRE~\cite{10341820} assumes fixed risky regions whose traversal costs depend on teammate support, while ~\cite{10341497} models deterministic team-state-dependent costs and ~\cite{duggan2025uncertaintyawareplanningheterogeneousrobot} extends this to probabilistic costs under environmental uncertainty. 
Support is coordinated through predefined spatial support relationships~\cite{10341820}, team configurations~\cite{10341497}, or information-gathering actions~\cite{duggan2025uncertaintyawareplanningheterogeneousrobot}.
However, these approaches do not explicitly model stochastic adversaries as a source of environmental uncertainty or capture how their motion changes traversal risk over time. 
Meanwhile, prior work on planning under dynamic obstacles or threats~\cite{10.1007/s10514-013-9334-3,aoude2010threat} predicts their future locations to avoid high-risk regions, but does not model teammate support as a cooperative mechanism for mitigating traversal cost in evolving risky regions.

We formulate multi-robot graph traversal problem with stochastically moving adversaries, while allowing the robot team to coordinate teammate support.
When adversaries remain stationary, our formulation reduces to the static risky-edge setting problem of TCGRE [1].
In the stochastic case, we model each adversary’s motion as a first-order Markov stay–move process over graph edges and propagate its occupancy distribution over a finite planning horizon to obtain time-indexed edge-risk forecasts.
We then use these forecasts to inform support candidate selection and joint path planning.
We show that forecast-informed support decisions consistently reduces expected team cost relative to the evaluated baselines in stochastic motion settings across different graph sizes and robot team configurations.



\section{Related Work}
\label{sec:related_work} 
Multi-agent path finding (MAPF) coordinates multiple agents on graphs, with classical methods ~\cite{sharon2015conflict,Silver2005Cooperative} primarily addressing agent-to-agent collision avoidance under fixed traversal costs. 
Uncertain-graph formulations include Canadian Traveler Problem (CTP)~\cite{bar1991canadian} and its variants for multi-agent settings~\cite{Shiri2017OnTheOnline, zhang2013kcanadian}, where edge availability or traversal conditions are uncertain and may be revealed during execution. 
Our setting likewise considers multi-robot graph traversal under environmental uncertainty, but does not impose MAPF-style collision constraints or blocked edge uncertainty as in CTP.
Instead, we consider dynamic environmental risk that changes traversal costs over time and can be mitigated through teammate support.

Time-dependent shortest-path and temporal-graph methods~\cite{Wang2019TimeDependentGD,DingYQ08,Dean2004} model time-varying edge costs, connectivity, or both, while stochastic shortest-path (SSP)~\cite{Bertsekas1991AnAO} and multi-agent SSP variants~\cite{jonáš2026multiagentstochasticshortestpath,Khonji2022MultiAgentCS}, together with robust MAPF formulations \cite{Khonji2022MultiAgentCS,atzmon2020robust, atzmon2020probabilistic} incorporate probabilistic costs, delays, or chance constraints. 
These works provide mechanisms for reasoning about temporal or uncertain graph conditions, but generally do not couple evolving environmental risk with cooperative support decisions that modify traversal costs. 

Dynamic-obstacle and threat-aware planners~\cite{10.1007/s10514-013-9334-3,aoude2010threat} predict future obstacle motion or threat exposure to find path feasibility or avoid collisions or high-risk regions during planning.
These works capture evolving environmental conditions, but do not model teammate support as a cooperative mechanism for mitigating traversal costs associated with those conditions.

Stochastic patrolling and surveillance methods \cite{Duan2020MarkovCS,ijcai2020p582,bosansky2011computing} employ randomized or Markov-based robot team strategies to reduce predictability and improve robustness against adversarial observation or exploitation.
Pursuit-evasion~\cite{decarufel2024copsrobberperiodic,Erlebach2024EdgePeriodic} and adversarial graph-traversal methods~\cite{berneburg2024multirobotcoordinationinducedadversarial} similarly often involve strategic opponents and two-sided optimization. 
In contrast, our problem considers goal-directed multi-robot traversal with one-sided optimization in which environmental entities follow stochastic motion inducing time-varying traversal costs.

The closest related work considers cooperative multi-robot graph traversal with state-dependent costs~\cite{10341820,10341497, duggan2025uncertaintyawareplanningheterogeneousrobot}. 
TCGRE~\cite{10341820} introduces fixed risky regions whose traversal costs can be reduced through teammate support from predefined spatial positions. 
~\cite{10341497} models deterministic team-state-dependent traversal costs that depend on team configuration such as overwatch and formations, while its uncertainty-aware extension~\cite{duggan2025uncertaintyawareplanningheterogeneousrobot} models probabilistic state-dependent costs under environmental uncertainty and uses scout robots to gather information to reduce this uncertainty.
These formulations establish multi-robot coordination under state-dependent or uncertain traversal costs, but do not explicitly model stochastic motion of individual risk-inducing entities or capture how their motion changes traversal risk over time. 

Our work generalizes support-based multi-robot graph traversal from fixed risky regions~\cite{10341820} to evolving risk induced by stochastically moving entities, where teammate support mitigates the resulting time-varying traversal costs.

\section{Problem Formulation}
\label{sec:problem} 
The formulation specifies robot and adversary dynamics, support model, and robot team objective on a temporal graph.

\subsection{Robots and Environment}
We consider a temporal graph $G=(V,E,\{c_t\}_{t=0}^T)$ with discrete time $t=0,1,\dots,T$, 
where each $c_t:E\to\mathbb{R}_{\ge0}$ assigns a time-dependent cost $c_t(u,v)=c((u,v),t)$ to every edge $(u,v)\in E$.
For each node $u\in V$, let $\mathcal{N}_V(u)$ denote the set of nodes $(v\in V\mid (u,v)\in E)$ adjacent to $u$.

There are $N$ robots indexed by $i=1,\dots,N$, each with a start node $s_i\in V$ and a goal node $g_i\in V$, who can stay in current node or move to adjacent node at each step t.  
Let $x_i(t)\in V$ denote the node occupied by robot $i$ at time $t$. At each time step $t$, robot $i$ selects an action $a_i(t)\in \{\mathrm{wait},\mathrm{move}, \mathrm{support}\}.$
The robot’s state then evolves as:
\[
x_i(t{+}1)=
\begin{cases}
x_i(t), & a_i(t)=\mathrm{wait}\ \text{or}\ a_i(t)=\mathrm{support}, \\[4pt]
v, & a_i(t)=\mathrm{move}(u,v),\ x_i(t)=u.
\end{cases}
\]
Both $\mathrm{wait}$ and $\mathrm{support}$ keep the robot at its current node.
A $\mathrm{support}$ action activates support for eligible edge traversals subject to support conditions defined in (Sec.~\ref{subsec-support-model}), whereas $\mathrm{wait}$ provides no such effect.

\subsection{Adversary Dynamics}
\label{subsec-adv-dynamics}
There are $M$ adversaries indexed by $j=1,\dots,M$, each occupying  edge $e=(u,v)\in E$ at any time. 
For an edge $e=(u,v)$, let $\mathcal{N}_E(e)$ denote the set of neighboring edges $(e'=(u,v') \mid e' \in E)$  that share a common node $u$ with $e$, excluding $e$ itself.

Let $e^j_t$ denote the random edge occupied by adversary $j$ at time $t$. 
At each step, an adversary either \emph{stays} on its current edge or \emph{moves} to an adjacent edge according to stochastic process~$\theta$:
\[
\Pr[e_{t+1}^j=e \mid e_t^j=e] \triangleq \theta_{\mathrm{stay}}(e),
\]
\[
\Pr[e_{t+1}^j=e' \mid e_t^j=e]
\triangleq \theta_{\mathrm{move}}(e\rightarrow e'),
\qquad e'\in\mathcal{N}_E(e).
\]

These probabilities satisfy the normalization constraint
\[
\theta_{\mathrm{stay}}(e)
+
\sum_{e'\in\mathcal{N}_E(e)}
\theta_{\mathrm{move}}(e\rightarrow e')
=1.
\]
Together, they define a first-order Markov chain over the edge set $E$, where transitions occur only between adjacent edges sharing a common node~\cite{norris97}.

In this work, we assume transitions are distributed uniformly across adjacent edges. Thus the probability becomes
\[
\theta_{\mathrm{move}}(e\rightarrow e')
=
\frac{1-\theta_{\mathrm{stay}}(e)}
{|\mathcal{N}_E(e)|},
\qquad
e'\in\mathcal{N}_E(e),
\]
 or zero otherwise.

For each adversary \(j\), let \(\gamma_{uv}^{A,j}(t)\in\{0,1\}\) indicate whether adversary \(j\) occupies edge \((u,v)\) at time \(t\). The total number of adversaries occupying the edge is
$$
n_{uv}^{A}(t)
\triangleq
\sum_{j=1}^{M}\gamma_{uv}^{A,j}(t).
$$

In this work, we assume edge risk depends only on whether at least one adversary occupies the edge. 
We define the aggregate adversary occupancy as binary indicator: 
$$
\gamma_{uv}^{A}(t)
\triangleq
\mathbb{I}\!\left[n_{uv}^{A}(t)\geq1\right]
\in\{0,1\}.
$$
The corresponding marginal edge-risk probability is
$$
\rho_{uv}(t)
\triangleq
\mathbb{E}\!\left[\gamma_{uv}^{A}(t)\right]
=
\Pr\!\left[\gamma_{uv}^{A}(t)=1\right]
\in[0,1].
$$
The computation and propagation of $\rho_{uv}(t)$ are described in Sec.~\ref{subsec:risk-forecast}.

\subsection{Support Model}
\label{subsec-support-model}
We define support as a coordinated team action in which robot occupy designated support positions and take a support action to mitigate the traversal risk experienced by its teammate traversing edges that are covered from those positions.

Let \(V_s\subseteq V\) denote the nodes that can serve as support positions, and \(H(x)\subseteq E\) denote the edges that can be supported from node \(x\) (e.g., due to sensing, communication, or coverage constraints). 
The number of supportable edges $|H(x)|$, may vary across support positions.
If $|H(x)|>1$, a robot located at $x$ can provide support coverage for all edges in ${H}(x)$ simultaneously.
For edge $(u,v) \in H(x)$, the set of feasible support locations is  $\mathcal{S}_{uv} = \{ x\in V_s : (u,v)\in H(x)\}$.
A robot may take the support action only at a support-capable node \(x\in V_s\). Thus, if \(x_i(t)\in V_s\), the robot may remain stationary by choosing either \(a_i(t)=\mathrm{wait}\) or \(a_i(t)=\mathrm{support}\); if \(x_i(t)\notin V_s\), remaining stationary corresponds only to \(a_i(t)=\mathrm{wait}\). 



We introduce a binary variable $\gamma^S_{uv}(t)\in\{0,1\}$, where $\gamma^S_{uv}(t)=1$ indicates that support is active for $(u,v)$ at time $t$.
For this work, $\gamma^S_{uv}(t)$ does not track which or how many robots provide support or receive, but only whether support is active on the edge $(u,v)$ i.e., at least one robot is positioned at a feasible support location $x\in S_{uv}$ and takes the support action while at least one teammate traverses $(u,v)$ during the same time step. Otherwise, $\gamma^S_{uv}(t)=0$.

A planner may optionally restrict the feasible relationships to a set of prioritized support candidates (e.g., due to resource constraints, computational efficiency). 
Such candidate set may be fixed a priori, selected once before planning, or updated dynamically during execution.
Let $\Gamma_{uv}(t) \subseteq \mathcal{S}_{uv}$ denote the candidate support locations considered for edge \((u,v)\). 
When no candidate restriction is imposed, all feasible support locations are available such that $\Gamma_{uv}(t) = \mathcal{S}_{uv}$. The forecast-informed candidate-selection strategy used in this work is described in Sec.~\ref{subsec:support-allocate}.

\subsection{Action costs}
Let $r_a>0$ denote the base traversal cost, and let $r_p>0$ represent the additional penalty incurred when traversing an adversary-occupied edge.
Given the adversary-occupancy indicator \(\gamma^A_{uv}(t)\) and support-activation indicator \(\gamma^S_{uv}(t)\), the traversal cost for edge \((u,v)\) at time \(t\) is
\begin{equation}
\label{eq:true-cost}
c_{\mathrm{move}}\big((u,v),t;\gamma^A,\gamma^S\big)
= r_a + r_p \, \gamma^A_{uv}(t)\,(1-\gamma^S_{uv}(t)).
\end{equation}

Waiting at a node incurs a constant cost:
\begin{equation}
\label{eq:wait-cost}
c_{\mathrm{wait}}(t)=w_c, \qquad w_c \ge 0,
\end{equation}

Supporting at a node incurs a constant cost:
\begin{equation}
\label{eq:support-cost}
c_{\mathrm{support}}(t)=w_s, \qquad w_s \ge 0.
\end{equation}

Upon arrival at $g_i$, the robot enters a goal state and incurs no further cost.
The per-robot action cost then becomes:
\begin{equation}
c^{(i)}_{\mathrm{act}}(t) =
\begin{cases}
c_{\mathrm{move}}\big((u,v),t;\gamma^A,\gamma^S\big), & a_i(t)=\mathrm{move}(u,v), \\[4pt]
c_{\mathrm{wait}}(t), & a_i(t)=\mathrm{wait}, \\[4pt]
c_{\mathrm{support}}(t), & a_i(t)=\mathrm{support}, \\[4pt]
0, & x_i(t)=g_i. \
\end{cases}
\end{equation}
\textit{Remark.}
Small positive costs $w_c, w_s > 0$ are used to prevent degenerate strategies (e.g., indefinite waiting or cost-free support behavior), while keeping traversal risk the primary contributor to total cost in our experiments.

\subsection{Team Objective}
We define the joint state for $N$ robots at time $t$ as $ s(t)=\left(x_1(t),\ldots,x_N(t),\gamma^A(t)\right)$, where $\gamma^A(t)$ denote the realized adversary occupancy on the edges defined in Sec.\ref{subsec-adv-dynamics}. 
A joint policy $\pi$ maps the joint state to a joint action such that
$\pi:s(t)\mapsto \left(a_1(t),\ldots,a_N(t) \right), $where each $a_i(t)\in\{\mathrm{move},\mathrm{wait},\mathrm{support}\}$.
Given the joint state and action, $\gamma^S(t)=\{\gamma^S_{uv}(t):(u,v)\in E\}$ denotes the resulting binary support activation according to the support model defined in Sec.~\ref{subsec-support-model}.

The team planning objective is:
\begin{equation}
\label{eq:team-objective}
J_{\mathrm{real}}^{*}
=
\min_{\pi}
\mathbb{E}
\left[
\sum_{t=0}^{T}
\sum_{i=1}^{N}
c_{\mathrm{act}}^{(i)}(t)
\right],
\end{equation}

Variants of this objective can also include risk-sensitive criteria (e.g., variance, chance constraints or CVaR), but in the base formulation we minimize expected cumulative cost.





\section{Method}
\label{sec:method} 
\subsection{Overview and Pipeline}
\begin{figure*}[t!]
    \centering
    \includegraphics[width=0.8\linewidth]{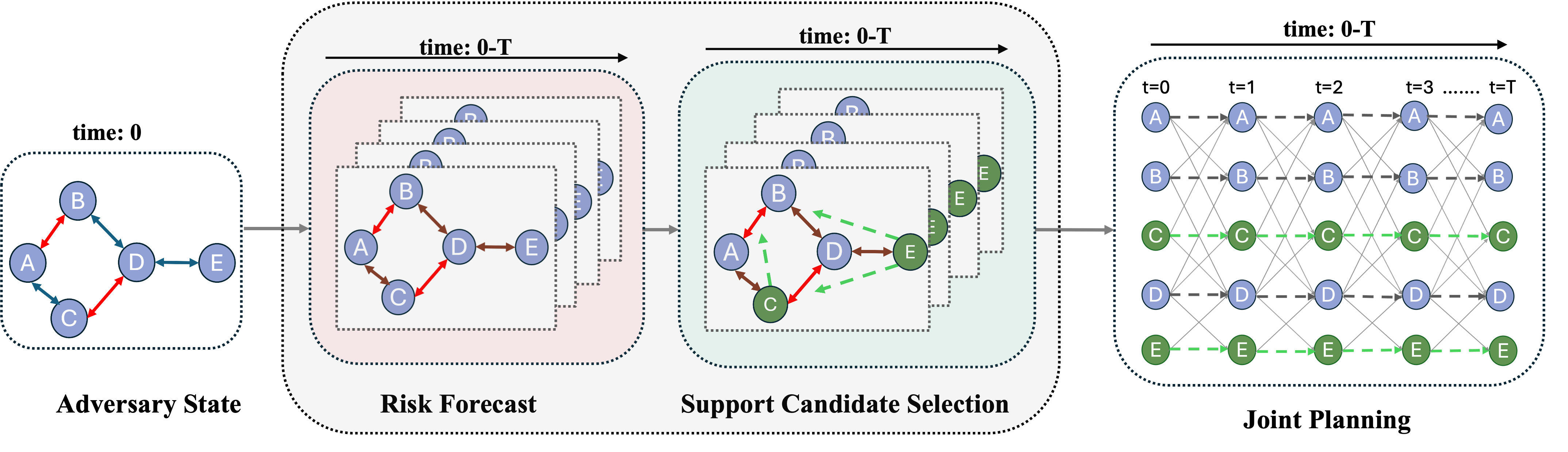}
 \caption{{\small  \textbf{Forecast-aware cooperative planning pipeline.} 
The graph (left) has adversaries with stochastic dynamics $\theta$. Their occupancy distributions are propagated forward in time to obtain edge-risk forecasts $\rho_{uv}(t)$ (red, middle). These forecasts guide support candidate selection, producing support-edge mappings $\Gamma_{uv}$ (green, middle). The resulting temporal graph (right) enables joint planning to minimize forecast-based expected team cost.}}

    \label{fig:pipeline}
\end{figure*}


Although the adversary transition dynamics are known (Sec.~\ref{subsec-adv-dynamics}), their future locations i.e., which edge each adversary will occupy at future timestep is uncertain. Accounting for all possible sequences of adversary edge occupancies during planning quickly becomes computationally expensive.
We therefore, propagate the known stochastic dynamics to obtain time-indexed edge-risk forecasts that inform both support candidate selection and joint path planning.
The proposed framework integrates three stages: i) adversarial risk forecasting, ii) forecast-informed support candidate selection, and iii) joint path planning on a temporal graph (Fig.~\ref{fig:pipeline}). 
The forecasts identify potentially risky edges (Sec. ~\ref{subsec:risk-forecast}) and prioritize feasible support locations using the heuristic scoring rule (Sec.~\ref{subsec:support-allocate}), after which joint planning minimizes expected traversal cost under the forecasted time-varying risk (Sec.~\ref{subsec:forecast-based-exp-cost}). 
Together, these stages enable anticipatory support coordination under known stochastic adversary motion.

\subsection{Adversary-Induced Edge Risk Forecast}
\label{subsec:risk-forecast}
Given the adversary stochastic dynamics defined in Sec.~\ref{subsec-adv-dynamics}, we propagate the occupancy distribution of each adversary forward in time and aggregate the resulting marginal probabilities to obtain time-indexed edge-risk forecasts $\rho_{uv}(t)$. 
Each adversary is modeled as a Markov process evolving over edges according to the stay-move transition model $\theta$. 
This evolution is captured by the edge-to-edge transition matrix $\Theta \in \mathbb{R}^{|E|\times|E|}$, with each entry $\Theta_{e',e}$ defined as:
\[
\Theta_{e',e} \;=\; 
\theta_{\mathrm{stay}}(e)\,\mathbb{I}[e'=e] 
+ \theta_{\mathrm{move}}(e\!\to\! e')\,\mathbb{I}[e'\sim e],
\]
where $\mathbb{I}[\cdot]$ is the indicator function and $e' \!\sim\! e$ 
denotes the two edges share a common node.

For each adversary $j \in \{1,2,..,M\}$, let $q^{(j)}_{uv}(t)\in[0,1]^{|E|}$ denote the marginal probability that adversary $j$ occupies edge $e=(u, v)$ at time $t$. 
This marginal distribution is propagated over the horizon $T$ as:
\[
q^{(j)}_{uv}(t+\tau) = (\Theta)^\tau q^{(j)}_{uv}(t),
\qquad \tau=0,1,\dots,T.
\]

The resulting edge-risk probability is obtained by combining the independent occupancy distributions of all $M$ adversaries:
\begin{equation}
    \rho_{uv}(t+\tau) \;=\; 1 - \prod_{j=1}^M \big(1 - q^{(j)}_{uv}(t+\tau)\big),
\end{equation}
Thus, $\rho_{uv}(t+\tau)$ represents the probability that at least one of the $M$ independent adversaries occupies edge $(u,v)$ at time $t+\tau$. This follows directly from the union probability of independent occupancy events~\cite{thrun2005probabilistic}.

\subsection{Forecast-aware Support Candidate Selection}
\label{subsec:support-allocate}
After forecasting edge risks $\rho_{uv}(t)$, we select prioritized support candidates from the feasible support locations defined in Sec. III-C. 
Unlike static approaches such as TCGRE [1], which consider fixed risky regions and support relationships, our method uses the forecasted risk evolution over horizon (T) to determine which feasible support locations are prioritized. 
Candidate selection is performed once at the initial planning stage $(t=0)$ where the selected candidate locations are not mandated support assignments, but eligible positions that the planner may use during search. 
The selection process consists of three phases.

\subsubsection{Candidate Support Nodes}
We restrict candidate selection to edges with nonzero forecasted risk over horizon $T$. Let
$
E_{\mathrm{risk}} =
\left\{
(u,v)\in E
\mid
\exists\, t\leq T
\text{ s.t. }
\rho_{uv}(t)>0
\right\}
$
denote the set of such edges. For each $(u,v)\in E_{\mathrm{risk}}$, we define the candidate support set
$
C_{uv}
=
\left\{
x\in S_{uv}
:
\min\big(d(x,u),d(x,v)\big)\leq k
\right\},
$
where $S_{uv}$ denotes the feasible support locations for edge $(u,v)$, as defined in Sec.~III-C, and $k$ specifies the neighborhood size. 
$k$ is an optional locality restriction used to reduce the candidate search space, where $k\leq |V|-1$.
Thus, $C_{uv}\subseteq S_{uv}$ restricts consideration to feasible support locations within $k$ hops of the corresponding edge.

\subsubsection{Node Scoring}

To prioritize candidate support locations, each $x\in C_{uv}$ is assigned a heuristic score that combines path overlap with forecasted risk:
\begin{equation}
\mathrm{Score}(x)
=
\alpha \hat{P}(x)
\left(
1+\beta \hat{R}(x)
\right).
\label{eq:node-score}
\end{equation}
where
\begin{itemize}
    \item$\hat{P}(x)$ (Normalized Path Overlap): Defined as $\hat{P}(x)=\dfrac{P(x)}{\max_{y\in V}P(y)}$ where $P(x)=\sum_{i=1}^N\mathbb{I}[x\in\text{path}_i^{T}]$ counts how many robots' shortest paths $\text{path}_i^{T}$ traverse $x$. 
    This ensures support is placed along high-traffic corridors.
      \item $\hat{R}(x)$ (Normalized Risk Potential): Defined as $\hat{R}(x)=\dfrac{e^{R(x)}}{\sum_{y\in C_{uv}}e^{R(y)}}$ as a value derived from the raw risk potential $R(x)$ where 
        \[
    R(x)=\frac{R_{uv}^{\text{tot}}}{1+\min(d(x,u),d(x,v))}, 
    \quad 
    R_{uv}^{\text{tot}}=\sum_{t=1}^{T}\rho_{uv}(t).
    \]
    This term prioritizes nodes that are spatially optimal for covering the cumulative risk expected on edge $(u,v)$.
    
    \item $\alpha,\beta$ (Weighting parameters): These parameters (default $\alpha = \beta = 1$) allow the system to tune the importance of traffic overlap vs. environmental risk.
\end{itemize}

\subsubsection{Selection Rule}
For each $(u,v)\in E_{\mathrm{risk}}$, we select the top-$s$
candidate locations according to their scores:
\[
\Gamma_{uv} = \operatorname{arg\,top}_{s}\big\{ \text{Score}(x): x \in C_{uv} \big\}.
\]

This yields a compact support-candidate mapping
$
\Gamma
=
\left\{
\Gamma_{uv}
\right\},
$
where each edge with forecasted risk is associated with up to $s$ prioritized support locations according to the forecast-aware scoring function (see Alg.~\ref{alg:support}). 
Because candidate selection is performed once at the initial planning stage $(t=0)$, the selected sets remains fixed over the planning horizon, i.e.,
$\Gamma_{uv}(t)=\Gamma_{uv},$ for all $t \in T$.
\begin{algorithm}[t]
\caption{Forecast-Aware Support Candidate Selection}
\label{alg:support}
\KwIn{Graph $G=(V,E,\{c_t\}_{t=0}^T)$, risk forecasts $\rho_{uv}(t)$, feasible support sets $S_{uv}$, robot paths $\{\text{path}_i\}_{i=1}^N$, horizon $T$,  neighborhood size $k$,    candidate supports per edge $s$}
\KwOut{Support mappings $\Gamma=\{\Gamma_{uv}\}$}
$E_{\mathrm{risk}} \gets \{(u,v)\in E:\exists t\le T,\rho_{uv}(t)>0\}$ \\
\For{$(u,v)\in E_{\mathrm{risk}}$}{
\hspace{0.3cm} $C_{uv} \gets \{x\in  S_{\mathrm{uv}} : \min(d(x,u),d(x,v))\le k\}$ \\
\hspace{0.3cm} \For{$x\in C_{uv}$}{
\hspace{0.6cm} compute $\text{Score}(x)$ using Eq.~\eqref{eq:node-score} } 
\hspace{0.3cm}$\Gamma_{uv} \gets 
        \operatorname{arg\,top}_{s}\{\,\text{Score}(x) : x \in C_{uv}\,\}$
}
\Return $\Gamma$
\end{algorithm}

\subsection{Forecast-based Team Objective}
\label{subsec:forecast-based-exp-cost}
Equ.~\ref{eq:team-objective} uses \(\gamma^A_{uv}(t)\in\{0,1\}\) to indicate whether edge \((u,v)\) is occupied by an adversary at time \(t\). 
We replace this occupancy indicator \(\gamma^A_{uv}(t)\) with its expected value, the edge-risk forecast \(\rho_{uv}(t)=\mathbb{E}[\gamma^A_{uv}(t)]\in[0,1]\). 
Under this expectation, the traversal cost for edge \((u,v)\) at time \(t\) is:

\begin{equation}
\label{eq:expected-cost}
\bar c_{\mathrm{move}}\big((u,v),t;\gamma^S\big)
= r_a + r_p\,\rho_{uv}(t)\,(1-\gamma^S_{uv}(t)).
\end{equation}

The per-robot expected action cost is:
\begin{equation}
\bar c^{(i)}_{\mathrm{act}}(t) =
\begin{cases}
\bar c_{\mathrm{move}}\big((u,v),t;\gamma^S\big), & a_i(t)=\mathrm{move}(u,v), \\[4pt]
c_{\mathrm{wait}}(t), & a_i(t)=\mathrm{wait}, \\[4pt]
c_{\mathrm{support}}(t), & a_i(t)=\mathrm{support}, \\[4pt]
0, & x_i(t)=g_i.
\end{cases}
\end{equation}

The approximated team planning objective is:
\begin{equation}
\label{eq:team-exp}
J_{\text{exp}} \approx \min_{\pi}\; 
\mathbb{E}\!\Bigg[\sum_{t=0}^{T}\sum_{i=1}^{N} \bar c^{(i)}_{\mathrm{act}}(t)\Bigg].
\end{equation}


\subsection{Planner}
Given the time-indexed edge-risk forecasts and selected support candidates, we perform centralized joint planning on the resulting temporal graph using a lazily evaluated A* search. 
Risk-dependent transition costs are evaluated only when the corresponding joint transitions are considered, while the search heuristic ignores risk and support effects. 
For joint state \(\tilde{s(t)}=(x_1(t),\ldots,x_N(t))\), the heuristic is $ h(\tilde{s(t)})=r_a\sum_{i=1}^{N} d(x_i(t),g_i)$, where \(d(x_i(t),g_i)\) is the shortest-path distance of robot \(i\) from its current position to its goal. 
Since every edge traversal costs at least \(r_a\) and wait and support costs are nonnegative, this heuristic is admissible and consistent.


\section{Experiments}
\label{sec:experiments} 

\begin{figure*}[t]
    \centering
    \includegraphics[width=0.8\textwidth]{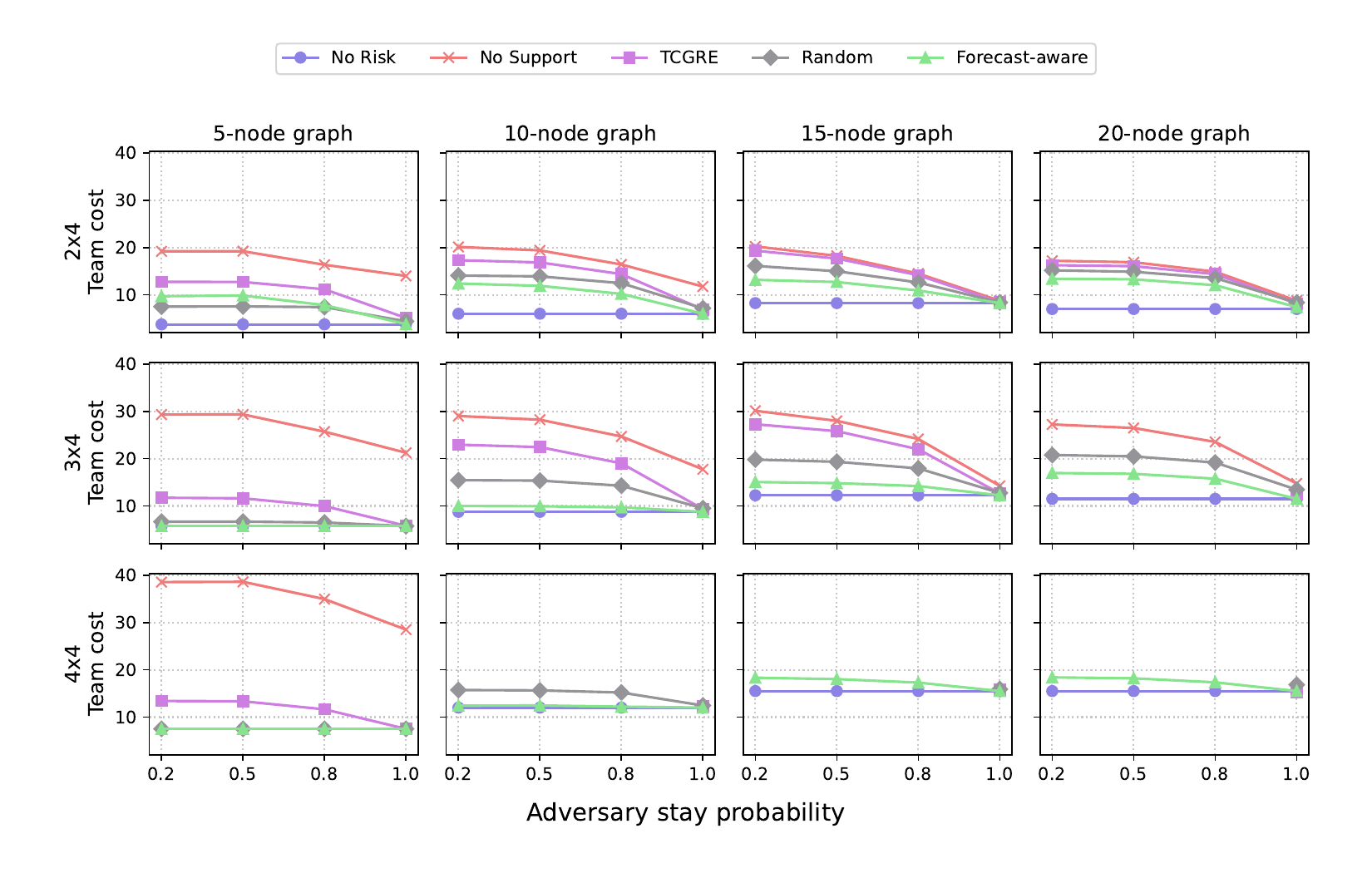}
    \caption{ {\small Expected team cost ($J_{\exp}$) across adversary stay probabilities $\{0.2, 0.5, 0.8, 1.0\}$, graph sizes ($|V| \in \{5, 10,15,20\}$), and robot--adversary configurations $\{ 2\times4, 3\times4, 4\times4 \}$. Rows show configurations and columns show graph sizes. }}
    \label{fig:expected-team-cost}
\end{figure*}


\subsection{Experimental Setup}

\textbf{Graph Environments.} 
We evaluate on randomly generated undirected graphs of $|V| \in \{5, 10, 15, 20\}$ nodes and edge-to-node ratios $r \in \{1.2, 1.4, 1.6, 1.8\}$. 


\textbf{Robots and Adversaries.}  
Each scenario contains 2, 3, or 4 robots with a fixed set of 4 adversaries. 
Robots move deterministically between nodes toward assigned goals. 
Adversaries follow a stochastic stay-move Markov process along edges, parameterized by stay probability. 
We evaluate four stay probabilities i.e., $stay \in \{0.2,0.5,0.8,1.0\}$ spanning from high to zero mobility.


\textbf{Task Configurations.}
We consider two start-goal settings: DSDG (Different Start-Different Goal) is used for all experiments, while SSSG (Same Start-Same Goal) is used only for ablations (\ref{ab:ablation2}).
Adversaries are randomly initialized such that no two adversaries occupy the same edge in both settings at $t=0$.

\textbf{Horizon.}  
The horizon $T$ is set proportional to the graph size ($T\propto|V|$) for both risk forecasting and planning, ensuring sufficient lookahead and planning time, with a 90-second of maximum runtime limit for all methods.

\textbf{Support Candidate Selection.} 
Support nodes are selected from the k-hop neighborhood ($k$=2) with at most one support node per risky edge ($s$=1) within the horizon $T$.

\textbf{Traversal Cost.} Each edge traversal incurs a base cost $r_a=1$. An adversarial encounter incurs penalty $r_p=10$. 

\textbf{Evaluation Protocol.}   
For each graph size and stay probability, four independent graph instances are evaluated; for each instance, five independent trials (distinct random seeds) are executed. Each (instance, seed) pair constitutes one run, yielding 4×5=20 runs per stay probability. Results are averaged over these runs. All experiments are conducted on a Mac M1 with 8 GB RAM.


\subsection{Baselines}
We compare our \emph{\textbf{Forecast-aware}} support candidate selection method against four baselines. 
\textbf{No Risk:} Adversarial risk and support nodes are disabled ($r_p=0, S_{uv}=\emptyset$), yielding time-invariant edge costs ($c_{\mathrm{move}}=r_a$) and an optimistic lower bound on team cost.
\textbf{No Support:} Forecasted risk is present but no support nodes selected ($r_p\not=0, S_{uv}=\emptyset$). 
\textbf{Random:} Forecasted risk is present and support nodes selected uniformly at random for edges that become risky within the forecast horizon. 
\textbf{TCGRE:} Forecasted risk is present and support nodes are selected using the static initial-snapshot rule of  TCGRE~\cite{10341820} i.e., only for edges that are risky at \(t=0\).
For all support candidate selection methods (\textbf{Random}, \textbf{TCGRE}, and \emph{\textbf{Forecast-aware}}) support nodes are selected from candidate nodes within the same $k$-hop neighborhood, ensuring differences arise solely from the candidate selection strategy.

\textbf{Monte Carlo Evaluation (Realized Performance).}
We use Monte Carlo (MC) simulation to sample adversary trajectories under the same stay-move dynamics $\theta$ and initial risk distribution $\rho_{uv}(0)$ to compute realized team cost  to calibrate against expected team cost induced by the risk forecast $\rho_{uv}(t)$ (Sec.~V-D.1).

\subsection{Main Results}
\subsubsection{Quantitative Results}
Across all graph sizes, robot-adversary configurations, and adversary stay probability settings shown in Fig.~\ref{fig:expected-team-cost}, the expected team cost decreases monotonically as the stay probability increases. 
The No Risk baseline consistently forms the lower bound, while No Support forms the upper bound. 
Our \emph{Forecast-aware} method consistently achieves the lowest expected team cost among all support candidate selection methods (\emph{Forecast-aware}, Random, and TCGRE) and remains closest to the No Risk lower bound for $stay = \{0.2, 0.5,0.8\}$ i.e., across all stochastic settings. Performance converge across all support candidate selection methods in the fully static case i.e., $stay = 1.0$, where temporal risk forecasting provides no additional benefit and the problem reduces to static-risk setting considered in TCGRE\cite{10341820}.

Wherever TCGRE produces feasible solutions, Random and \emph{Forecast-aware}  consistently outperform it because TCGRE select support candidates only for edges risky in initial snapshot, whereas Random and \emph{Forecast-aware} consider edges that become risky over the horizon. 
On the 5-node graph, the candidate space covers most of the graph, and many candidate support locations lie close to the robot routes. 
As a result, Random often selects a useful support location and performs comparably to \emph{Forecast-aware}. 
On larger graphs (\(|V|>5\)), uninformed random selection is less likely to choose useful candidates, while \emph{Forecast-aware} benefits from ranking candidates based on path relevance, proximity and forecasted risks.

\subsubsection{Qualitative Results}
\begin{figure*}[t]
    \centering
    \includegraphics[width=0.85\textwidth]{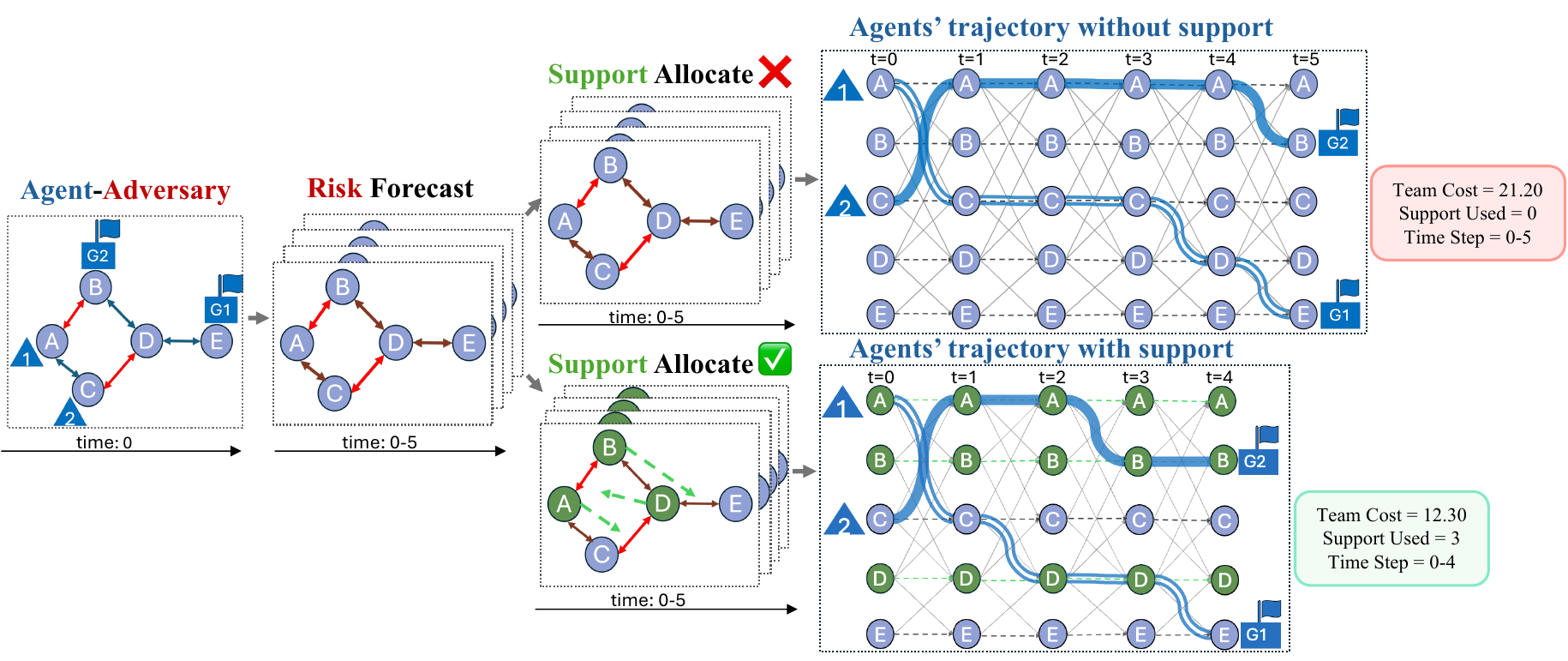} 
        \caption{{\small Illustrative example on a 5-node graph (stay = 0.8) comparing \emph{No Support} and the \emph{Forecast-aware} method.}}
    \label{fig:illus_example}
\end{figure*}

Fig.~\ref{fig:illus_example} illustrates 5-node graph example with two robots (blue) and two adversaries (red, $stay=0.8$) comparing \emph{No Support} with our \emph{Forecast-aware} method, under the same risk forecast and planning horizon ($T=5$). 
Under \emph{No Support}, robots respond to the high forecasted risk by waiting near risky corridors when near-term traversal is costly. 
After an initial position exchange, Robot~1 waits repeatedly at node $C$ before committing to $D$ $(A, C, C, C, D, E)$, while Robot~2 delays at $A$ before moving to $B$ $(C, A, A, A, A, B)$, resulting in higher cumulative cost and delayed goal arrival.

In contrast, \emph{Forecast-aware} selects support candidate positions in advance using the same time-indexed risk forecast, enabling earlier traversal of risky edges. 
After the same initial position exchange, Robot~1 follows $(A, C, D, D, E)$ while Robot~2 follows $(C, A, A, B, B)$, with coordinated support activated at key stages: Robot~2 supports Robot~1 from node $A$ during traversal of $(C, D)$, Robot~1 supports Robot~2 from node $D$ during traversal of $(A, B)$, and Robot~2 again supports Robot~1 from node $B$ during traversal of $(D, E)$. 
By leveraging forecast-informed support positions, robots avoid prolonged waiting and traverse risky corridors earlier, reducing team cost by approx. 42\% (21.20 to 12.30) and shortening completion time by one step (5 to 4).

\subsection{Ablations}
\label{ablation}
We further evaluate \emph{Forecast-aware} method through two ablation studies.
\subsubsection{Calibration of Forecast-based Expected Cost}
\label{acuracy_ablation}
To evaluate how well the forecast-based expected cost \(J_{\mathrm{exp}}\) (Eq.~\ref{eq:team-exp}) reflects realized cost \(J_{\mathrm{real}}\) (Eq.~\ref{eq:team-objective}), we compare \(J_{\mathrm{exp}}\) with the \(J_{\mathrm{real}}\) under MC-sampled adversary trajectories in Table I.
For each setting, \(J_{\mathrm{real}}\) is averaged over 2,500 realizations (500 trials for each of five random seeds).
Adversary initial locations are randomly initialized once and then fixed for the graph, while robot start-goal pairs are randomized for each robot-adversary configuration.
Across all settings, $J_{\text{real}}$ closely matches $J_{\text{exp}}$, with bounded deviations ($|\Delta| \le 1.0$). 
In most cases, $\Delta = J_{\text{real}} - J_{\text{exp}}$ is slightly negative.
This occurs because \(J_{\mathrm{exp}}\) uses forecasted occupancy probabilities that consider multiple possible future locations for each adversary rather than a single realized trajectory.
In contrast, \(J_{\mathrm{real}}\) is computed from sampled trajectories in which each adversary occupies only one specific edge at each time step. 
As a result, $J_{\text{exp}}$ can slightly overestimate the total traversal cost along the planned paths relative to a sampled realization.
In high-mobility settings i.e., $stay = 0.2$, with increasing number of adversaries $(Adv \ge 8)$ and robots $(Ag = \{3,4\})$, the increased stochasticity of adversary motion introduces higher variance in realized trajectories, leading to occasional small positive deviations.
These deviations remain consistently small and bounded, indicating that the forecast-based expected cost provides very close estimate of realized team cost.

\begin{table}[t]
\centering
\caption{{\small Cost calibration of expected vs.\ realized team cost ($J_{\text{exp}}$ vs $J_{\text{real}}$) across  stay probabilities $(stay = \{0.2, 0.5, 0.8\})$, robots $(Ag = \{2, 3, 4\})$, and adversaries $(Adv = \{2, 4, 6, 8\})$ on a 10-node graph ($r=1.6$, $k=2$, and $s=1$). Each \(J_{\mathrm{real}}\) averages 2,500 realizations (500 MC trials for each of five random seeds)}}
\label{tab:ablation_jexp_jreal}
\scriptsize
\renewcommand{\arraystretch}{1.0}
\setlength{\tabcolsep}{2.5pt}

\begin{tabular}{cc|c c c|c c c|c c c}
\toprule
\multirow{2}{*}{Ag} & \multirow{2}{*}{Adv} & 
\multicolumn{3}{c|}{stay = 0.2} &
\multicolumn{3}{c|}{stay = 0.5} &
\multicolumn{3}{c}{stay = 0.8} \\
\cmidrule(lr){3-5} \cmidrule(lr){6-8} \cmidrule(lr){9-11}
& &
$J_{\exp}$ & $J_{\text{real}}$ & $\Delta$ &
$J_{\exp}$ & $J_{\text{real}}$ & $\Delta$ &
$J_{\exp}$ & $J_{\text{real}}$ & $\Delta$ \\
\midrule

2 & 2 & 6.63 & 6.52 & \textbf{\textcolor{teal}{-0.11}} & 6.41 & 6.16 & \textbf{\textcolor{teal}{-0.25}} & 5.65 & 5.34 & \textbf{\textcolor{teal}{-0.31}} \\

2 & 4 & 8.18 & 8.06 & \textbf{\textcolor{teal}{-0.12}} & 8.06 & 7.93 & \textbf{\textcolor{teal}{-0.13}} & 7.09 & 6.62 & \textbf{\textcolor{teal}{-0.47}} \\

2 & 6 & 9.48 & 9.30 & \textbf{\textcolor{teal}{-0.18}} & 9.21 & 9.03 & \textbf{\textcolor{teal}{-0.18}} & 7.88 & 7.40 & \textbf{\textcolor{teal}{-0.48}} \\

2 & 8 & 10.77 & 10.71 & \textbf{\textcolor{teal}{-0.06}} & 10.69 & 10.52 & \textbf{\textcolor{teal}{-0.17}} & 9.43 & 8.56 & \textbf{\textcolor{teal}{-0.87}} \\
\midrule

3 & 2 & 9.61 & 9.30 & \textbf{\textcolor{teal}{-0.31}} & 9.36 & 9.12 & \textbf{\textcolor{teal}{-0.24}} & 8.74 & 8.52 & \textbf{\textcolor{teal}{-0.22}} \\

3 & 4 & 11.28 & 11.06 & \textbf{\textcolor{teal}{-0.22}} & 11.29 & 11.09 & \textbf{\textcolor{teal}{-0.20}} & 10.73 & 10.33 & \textbf{\textcolor{teal}{-0.40}} \\

3 & 6 & 12.64 & 12.55 & \textbf{\textcolor{teal}{-0.09}} & 12.55 & 12.31 & \textbf{\textcolor{teal}{-0.24}} & 11.69 & 11.03 & \textbf{\textcolor{teal}{-0.66}} \\

3 & 8 & 13.61 & 14.19 & \textbf{\textcolor{red}{+0.58}} & 13.61 & 13.41 & \textbf{\textcolor{teal}{-0.20}} & 12.80 & 12.16 & \textbf{\textcolor{teal}{-0.64}} \\
\midrule

4 & 2 & 11.30 & 11.06 & \textbf{\textcolor{teal}{-0.24}} & 11.06 & 10.91 & \textbf{\textcolor{teal}{-0.15}} & 10.49 & 10.41 & \textbf{\textcolor{teal}{-0.78}} \\

4 & 4 & 12.29 & 11.99 & \textbf{\textcolor{teal}{-0.30}} & 12.10 & 11.86 & \textbf{\textcolor{teal}{-0.24}} & 11.61 & 11.46 & \textbf{\textcolor{teal}{-0.15}} \\

4 & 6 & 13.10 & 12.63 & \textbf{\textcolor{teal}{-0.47}} & 12.75 & 12.44 & \textbf{\textcolor{teal}{-0.31}} & 12.08 & 11.82 & \textbf{\textcolor{teal}{-0.26}} \\

4 & 8 & 14.11 & 14.18 & \textbf{\textcolor{red}{+0.07}} & 13.09 & 13.52 & \textbf{\textcolor{teal}{-0.37}} & 12.97 & 12.52 & \textbf{\textcolor{teal}{-0.45}} \\
\bottomrule
\end{tabular}
\end{table}

\subsubsection{Effect of Scoring Factors for Candidate Support Selection}
\label{ab:ablation2}
\begin{figure}[t]
    \centering
    \includegraphics[width=\linewidth]{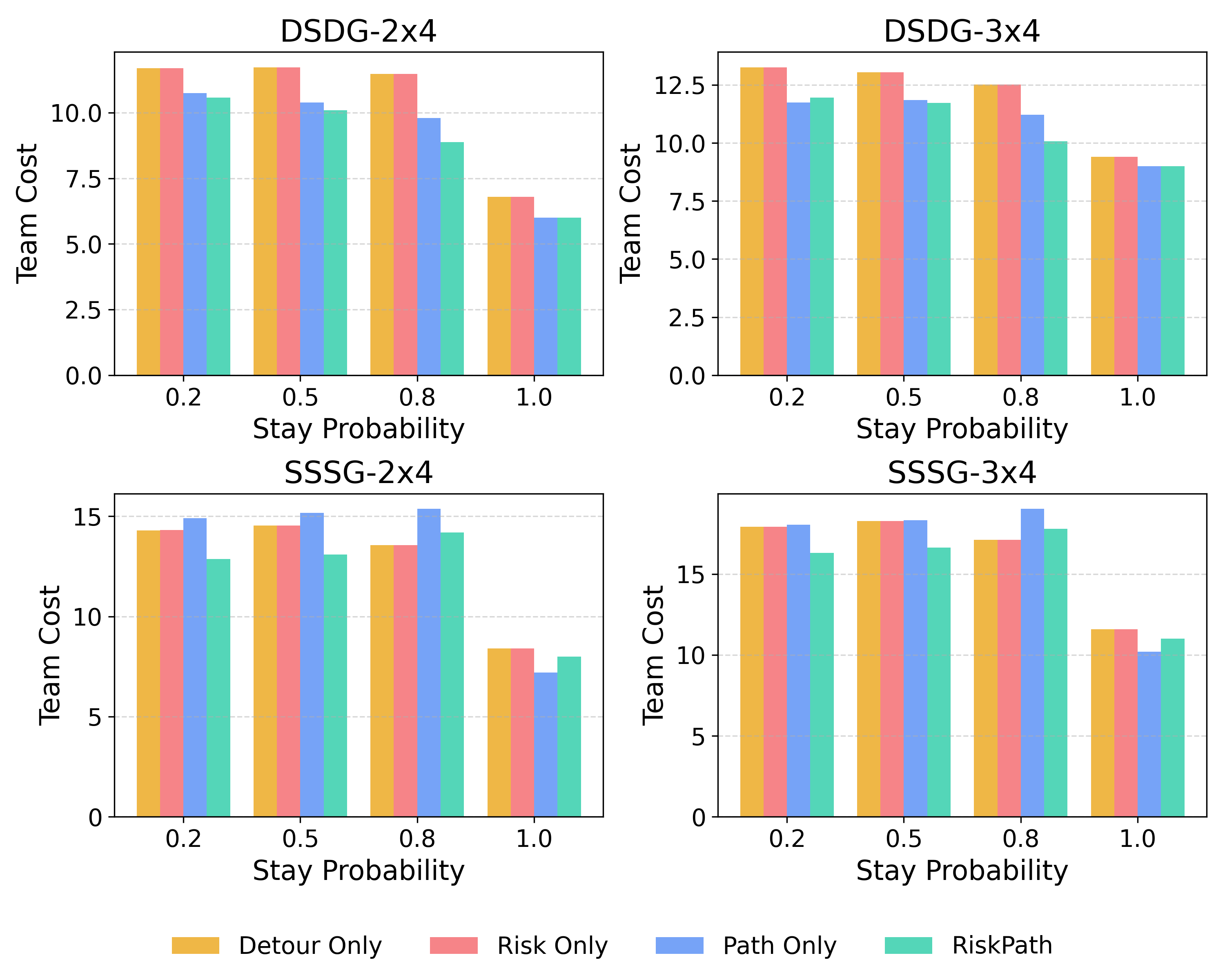}
    \caption{{\small Comparison of node-scoring factors across stay probabilities 
    $(stay = \{0.2, 0.5, 0.8, 1.0\})$ for DSDG and SSSG scenarios with $2 \times 4$ and $3 \times 4$ robot-adversary configurations on a 10-node graph ($r = 1.6$, $k = 2$, $s = 1$).}}
    \label{fig:ablation_node_score_support_barplots}
\end{figure}

Fig.~\ref{fig:ablation_node_score_support_barplots} compares four scoring strategies: RiskPath (risk forecast + path overlap), RiskOnly (risk forecast), PathOnly (path overlap), and DetourOnly (proximity) with RiskPath (Eq.~\ref{eq:node-score}) used as the default in the main experiments.
For this ablation, the same graph setup is used as in Ablation~\ref{acuracy_ablation} across stay probability settings, with both DSDG and SSSG task configurations.
Across DSDG scenarios, RiskPath consistently achieves the lowest team cost for $stay=\{0.2, 0.5, 0.8\}$, demonstrating the benefit of combining forecasted risk with anticipated robot traffic. 
At $stay = 1.0$, PathOnly performs comparably as adversarial risk becomes static and traffic flow becomes more informative for support selection. 
Minor deviations occur under high mobility (e.g., 3×4 at $stay = 0.2$), where PathOnly occasionally performs better suggesting that high stochasticity of adversary motion reduces the usefulness of temporal risk forecasts.
In SSSG scenarios, RiskPath performs best at higher mobility $(stay = \{0.2, 0.5\})$, while RiskOnly and DetourOnly become more competitive as mobility decreases $(stay = 0.8)$, indicating the increased importance of localized risk and proximity. 
At $stay = 1.0$, PathOnly yields the lowest cost, reflecting the dominance of robot traffic in the static case.
These results shows that combining temporal risk forecasts with robot traffic information is most effective in the general DSDG setting (Fig.~\ref{fig:expected-team-cost}).


\section{Conclusion}
We formulate multi-robot graph traversal problem with stochastically moving adversaries, where robots can utilize teammate support to mitigate high traversal costs induced by these adversaries. 
We propose a forecast-aware framework that predicts the temporal evolution of adversarial risk, prioritizes support candidate positions, and performs joint path planning on the resulting temporal graph.
This enables the robot team to anticipate evolving risk and effectively coordinate support when traversing high-risk regions.

The main experiments show that in stochastic motion settings $(stay=\{0.2,0.5,0.8\})$, \emph{Forecast-aware} consistently achieves lower expected team cost than the No Support upper bound and the Random and TCGRE support candidate-selection baselines while remaining closer to the No Risk lower bound.
Ablation studies further demonstrate that forecast-based expected cost closely matches realized cost under MC-sampled adversary trajectories and that integrating temporal risk forecasts with robot traffic information improves forecast-informed support candidate selection.
In the fully static case, $(stay=1.0)$, temporal forecasting provides no additional benefit and the problem reduces to the static-risk setting considered in TCGRE~\cite{10341820}.

Current limitations include reliance on known adversary dynamics and centralized planning, which may restrict applicability when transition models are unknown or when coordination must be performed in a decentralized manner. 
Future work will investigate online estimation of adversary dynamics, decentralized coordination, strategic adversaries that adapt their motion in response to the robot team, and larger-scale problems with longer planning horizons.


\section*{ACKNOWLEDGEMENT}
This work has taken place in the RobotiXX Laboratory at George Mason University. 
RobotiXX research is supported by National Science Foundation (NSF, 2350352), Army Research Office (ARO, W911NF2320004, W911NF2420027, W911NF2520011), Air Force Research Laboratory (AFRL), US Air Forces Central (AFCENT), Google DeepMind (GDM), Clearpath Robotics, Raytheon Technologies (RTX), Tangenta, Mason Innovation Exchange (MIX), and Walmart.

\section*{APPENDIX}
\subsection{Feasibility Results}
\begin{figure*}[t]
    \centering
    \includegraphics[width=0.8\textwidth]{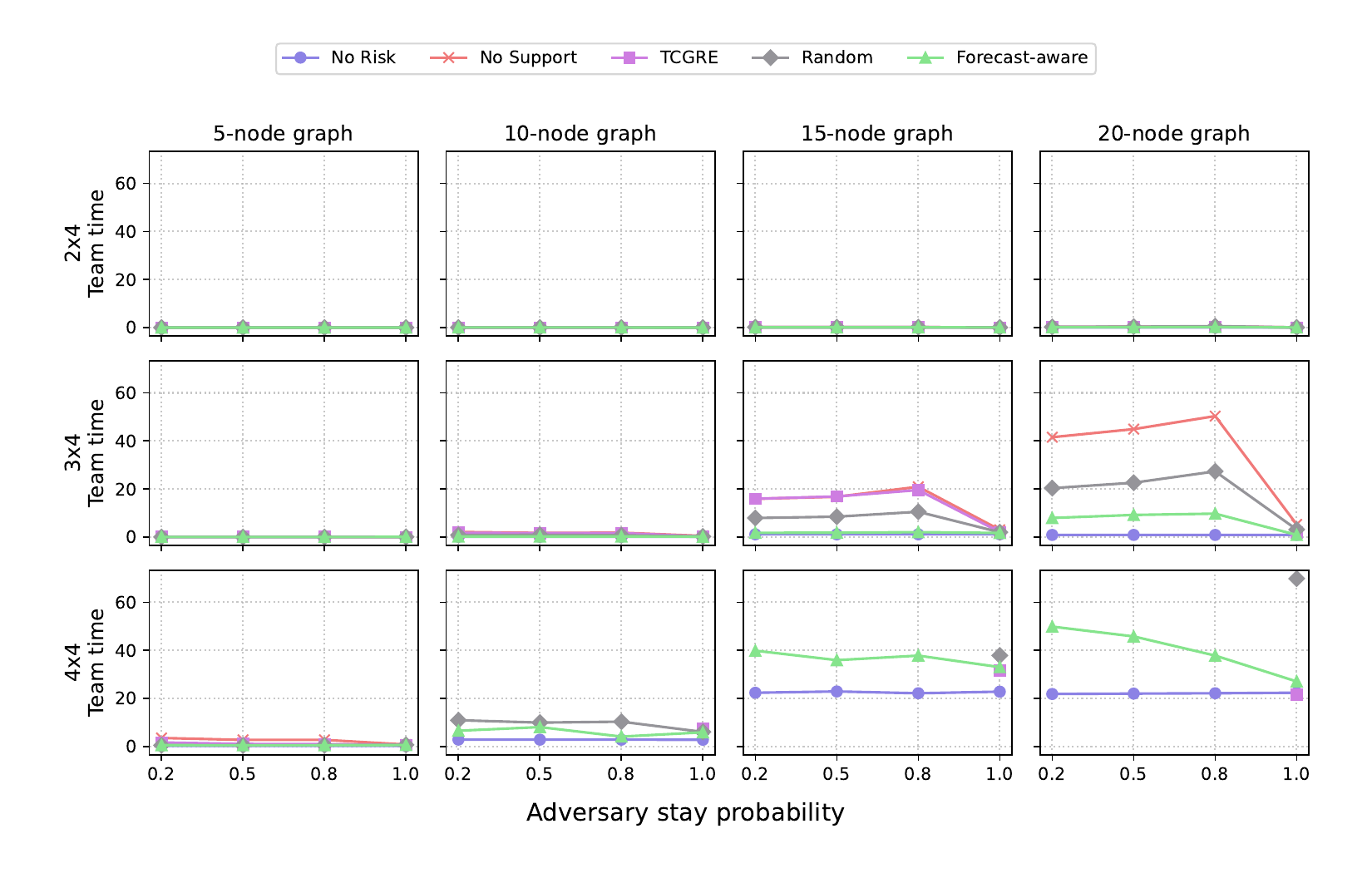}
    \caption{ {\small Average time taken corresponding to expected team cost ($J_{\exp}$) in Fig.~\ref{fig:expected-team-cost} across adversary stay probabilities $\{0.2, 0.5, 0.8, 1.0\}$, graph sizes ($|V| \in \{5, 10,15,20\}$), and robot--adversary configurations $\{ 2\times4, 3\times4, 4\times4 \}$. Rows show configurations and columns show graph sizes. }}
    \label{fig:avg_team_time}
\end{figure*}

We report average runtime statistics for all methods  under a fixed timeout of 90 seconds (Fig.~\ref{fig:avg_team_time}), corresponding to the computation of average expected team cost in Fig.~\ref{fig:expected-team-cost}. 
Across all graph sizes, agent-adversary configurations, and adversary stay probabilities, the \emph{Forecast-aware} method consistently terminated within the given timeout. 
By prioritizing support nodes aligned with predicted robot traffic, our method effectively prunes the search space, maintaining computational efficiency even as problem difficulty increases.

The No Risk lower bound baseline consistently achieves the lowest planning time, since it doesn't account for adversarial risk and support.
While the No Support upper bound baseline triggers timeouts in larger settings, such as 4-agent teams on 10-, 15-, and 20-node graphs.
This computational failure stems from the planner’s inability to find valid, low-risk trajectories without support coordination, leading to exhaustive searches beyond the time limit. 
\emph{Forecast-aware} is the only method that consistently achieves the lowest runtime in challenging regimes with high-agent settings (3$\times$4, 4$\times$4) on 10-, 15-, and 20-node graphs.
Meanwhile, methods such as TCGRE and Random frequently exceed the timeout in these setting due to insufficient or poorly placed support, which prevents the discovery of viable and timely paths to goals.

\bibliographystyle{ieeetr}
\bibliography{bib}

\end{document}